\documentclass[12pt,pra,aps,amssymb,amsmath,tightenlines,showpacs]{revtex4}
\usepackage{graphicx}
\usepackage{dsfont}

\newcommand{\half}{\mbox{$\textstyle \frac{1}{2}$}}
\newcommand{\quat}{\mbox{$\textstyle \frac{1}{4}$}}

\newcommand{\sqrtwo}{\mbox{$\textstyle \frac{1}{\sqrt{2}}$}}
\newcommand{\thetwo}{\mbox{$\textstyle \frac{\theta}{2}$}}

\newcommand{\rd}{\mbox{$\rm d$}}
\newcommand{\bea}{\begin{eqnarray}}
\newcommand{\eea}{\end{eqnarray}}

\begin{document}

\title{Dynamical state reduction in an EPR experiment}

\author{Daniel~J.~Bedingham}

\affiliation{Blackett~Laboratory,~Imperial College,~London~SW7~2BZ,~UK.}

\date{\today}

\begin{abstract}
A model is developed to describe state reduction in an EPR experiment as a continuous, 
relativistically-invariant, dynamical process. The system under consideration consists 
of two entangled isospin particles each of which undergo isospin measurements at spacelike 
separated locations. The equations of motion take the form of stochastic differential 
equations. These equations are solved explicitly in terms of random variables with 
{\it a priori} known probability distribution in the physical probability measure. 
In the course of solving these equations a correspondence is made between the state 
reduction process and the problem of classical nonlinear filtering. It is shown that 
the solution is covariant, violates Bell inequalities, and does not permit superluminal 
signaling. It is demonstrated that the model is not governed by the Free Will Theorem 
and it is argued that the claims of Conway and Kochen, that there can be no relativistic theory 
providing a mechanism for state reduction, are false.

\end{abstract}

\pacs{03.65.Ta, 03.65.Ud, 02.50.Ey, 02.50.Cw}

\maketitle

\section{Introduction}

The motivation for attempting to formulate a dynamical description of state reduction 
\cite{pearlorig0, pearlorig, gisin, ghir3, dios2, ghir2} stems from the
inherent problems of quantum measurement. In standard quantum theory the state reduction postulate is a 
necessary supplement to the Schr\"odinger dynamics in order that we can realise definite measurement 
outcomes from the potentiality of the initial state vector. The problem with this picture is that the 
pragmatic application of these two different laws of evolution is left to the judgment of the physicist 
rather than being fixed by exact mathematical formulation. Our experience in the use of quantum theory 
tells us that the state reduction postulate should not be applied to a microscopic system consisting of 
a few elementary particles until it interacts with a macroscopic object such as a measuring device. 
This works perfectly well in practice for current experimental technologies, but as we begin to explore systems 
on intermediate scales it is not clear whether state reduction should be assumed or not. A solution of the 
problem of measurement thus requires that we somehow set a fundamental scale to demarcate micro and macro 
effects within the dynamical framework. 

The formulation of an empirical model, objectively describing the dynamics of the state reduction
process is a direct approach to achieving this aim. 
The basic requirements we have for such a model can be characterised as follows \cite{Bass, Pear2}: 
\begin{itemize}
\item Measurements involving macroscopic instruments should have definite outcomes.
\item The statistical connections between measurement outcomes and the state vector 
prior to measurement should be preserved.
\item The model should be consistent with known experimental results.
\end{itemize}
The task of meeting these objectives in a relativistic context has met with technical difficulties
related to renormalization \cite{pear3, ghir1, adle, Pear, Nicr, tumul, pearshape, me}. 
These issues derive from the quantum field theoretic nature of
relativistic systems. In this paper we will attempt to sidestep this problem by considering a 
simplified quantum system with a finite-dimensional Hilbert space free from the problem of divergences.
Our aim is to elucidate the dynamical process of state reduction in a relativistic context.

We will consider a model describing the famous experiment devised by Einstein, Podolski, and 
Rosen (EPR) \cite{EPR}. The experiment involves two elementary particles in an entangled state and separated 
by a spacelike interval. The original purpose of EPR was to argue that quantum mechanics is fundamentally 
incomplete as a theory. In order to do this they made a locality assumption stating that the two particles 
are not able to instantaneously influence each other at a distance. 
Theoretical and experimental advances \cite{Bell,aspe} have since demonstrated the remarkable conclusion that the 
assumption of locality is incorrect. Entangled quantum systems can indeed transmit instantaneous 
influence at a distance when a measurement is performed. Although this fact negates the EPR argument, 
instead it poses questions for our understanding of quantum measurement. In particular, the notion of 
instantaneous influence due to state reduction during measurement seems to sit uncomfortably with the 
theory of relativity. 

A formal relativistically-covariant description of the state reduction associated with measurement 
has been given by Aharanov and Albert \cite{aa3}. They show that for a consistent description of the measurement 
process, the state evolution cannot take the form of a function on spacetime. The proposed solution 
is that state evolution should be described by a functional on the set of spacelike hypersurfaces as 
conceived by Tomonaga and Schwinger. This sets the scene for understanding how to formulate a fully 
dynamical and relativistic description of the state reduction process. 

Relativistic dynamical reduction models have been critically investigated from the perspective of the 
analysis of Aharanov and Albert by Ghirardi \cite{Ghir4}. There, the conceptual features of these models are 
discussed and shown to lead to a coherent picture. It is the intention of this work to extend the 
analysis of Ghirardi by constructing an explicit model of continuous state evolution. Our model,
which is described in detail in section \ref{sec:model}, is designed to highlight the peculiar nonlocal 
features. In sections \ref{sec:qsol} and \ref{sec:psol} we derive closed-form solutions to the 
stochastic equations of motion. The value of this is that it enables us to examine the nonlocal 
character of the stochastic noise processes. In section \ref{sec:filter} we apply the method of Brody 
and Hughston \cite{Dorj,Dorj2} to demonstrate that the equations describing the dynamical state reduction can be 
viewed as a description of a classical filtering problem. In section \ref{sec:bell} we generalise our 
model to consider an experiment where the experimenter can freely choose which measurement to 
perform on the individual particle from an incompatible set of possible measurements. This leads us 
to a discussion of the so-called Free Will Theorem \cite{fw1,fw2, Bass2, fwt2, fw3} of Conway and Kochen in section \ref{sec:freewill}. 
We use our findings to argue that the axiomatic assumptions of the Free Will Theorem are too restrictive
and that the conclusions of the theorem cannot be applied to dynamical models of state reduction.

\section{The model}
\label{sec:model}

We consider two particles denoted ${\it 1}$ and ${\it 2}$, each described by an internal isospin-$\half$ 
degree of freedom. The choice of an isospin system avoids complication encountered when dealing with 
conventional spin in a covariant formulation. The initial isospin state of the two particles is defined 
in spacetime on an initial spacelike hypersurface $\sigma_i$ as the isospin singlet state
\bea
|\psi(\sigma_i)\rangle = \sqrtwo\left\{
|+\half ; -\half\rangle - |-\half ; +\half\rangle
\right\}.
\eea
The isospin states for each particle are represented with respect to a fixed axis in isospin space. 

The particle trajectories in spacetime are assumed to behave classically. The two particles move in separate 
directions away from some specific location where they have been prepared. Each particle path
eventually intersects with the path of an isospin measuring device. This leads to a localised 
interaction which we assume takes place in some finite region of spacetime. We assume that the 
classical trajectories of the particles and measuring devices, and the finite regions of 
interaction are determined. Further we assume that the two measurement regions are completely
spacelike separated in the sense that every point in each region is spacelike separated from 
every point in the other region. We denote the two measurement regions by $R_{\it 1}$ and $R_{\it 2}$
(see figure 1).

\begin{figure}[t]
\begin{center}
\includegraphics[width=8cm]{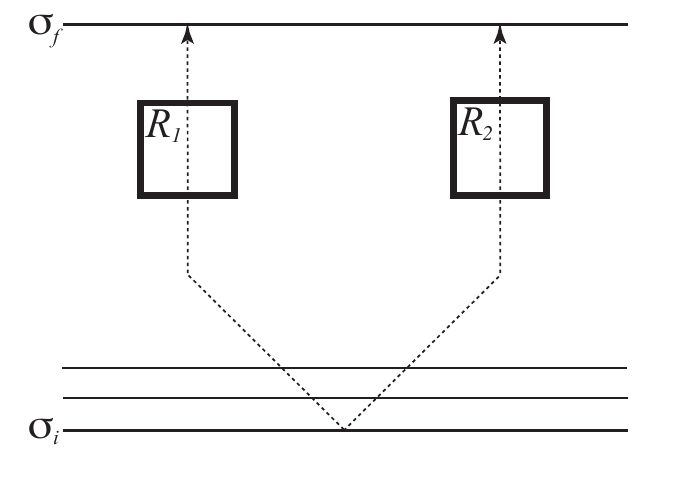}
\caption{\textsf{The diagram represents an experiment to measure the states of two entangled 
particles. The dashed lines are the (classical) particle trajectories where particle {\it 1} moves 
to the left and particle {\it 2} moves to the right. The vertical represents a timelike direction 
whilst the horizontal represents a spacelike direction. We suppose that within the spacetime 
region $R_{\it 1}$, a measurement is performed on particle ${\it 1}$. Similarly within the spacetime region 
$R_{\it 2}$ (spacelike separated from $R_{\it 1}$), a measurement is performed on particle ${\it 2}$. The initial 
state is defined on the spacelike hypersurface $\sigma_i$. The state advances as described by 
the Tomonaga picture through a sequence of spacelike surfaces defining a foliation of spacetime. 
}}
\end{center}
\end{figure}

In order to describe the state evolution we use the Tomonaga picture \cite{Tomo, Schw}. Standard unitary dynamics are 
described in this picture by the Tomonaga equation,
\bea
\frac{\delta|\psi(\sigma)\rangle}{\delta\sigma(x)} = -i H_{\rm int}(x)|\psi(\sigma)\rangle,
\label{TOM}
\eea  
where $H_{\rm int}$ is the interaction Hamiltonian.
Given two spacelike hypersurfaces $\sigma$ and $\sigma'$ differing only by some small
spacetime volume $\Delta \omega$ about some spacetime point $x$, the functional derivative is defined by
\bea
\frac{\delta |\psi(\sigma)\rangle}{\delta\sigma(x)} = \lim_{\sigma'\rightarrow\sigma}
\frac{|\psi(\sigma')\rangle-|\psi(\sigma)\rangle}{\Delta \omega}.
\eea
The operator $H_{\rm int}$ must be a scalar in order that equation (\ref{TOM}) has Lorentz invariant 
form. We must also have $[H_{\rm int}(x),H_{\rm int}(x')]=0$ for spacelike separated $x$ and $x'$ 
reflecting the fact that there is no temporal ordering between spacelike separated points. 

In differential form equation (\ref{TOM}) can be written
\bea
\rd_x |\psi(\sigma)\rangle = -i H_{\rm int}(x) \rd \omega |\psi(\sigma)\rangle
\label{TOMdif}
\eea 
where $\rd_x |\psi(\sigma)\rangle$ represents the infinitesimally small change in the state as the 
hypersurface $\sigma$ is deformed in a timelike direction at point $x$.

\begin{figure}[t]
\begin{center}
\includegraphics[width=8cm]{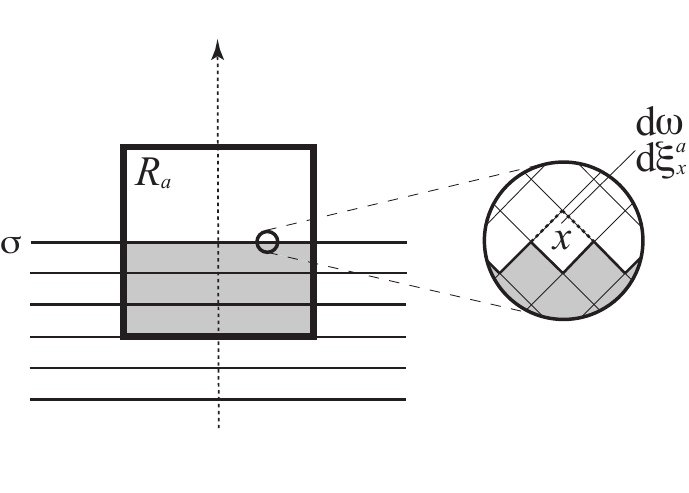}
\caption{\textsf{ The diagram represents a sequence of spacelike hypersurfaces advancing through the spacetime 
region $R_a$. The gray shading within $R_a$ corresponds to the spacetime volume $\omega_{\sigma}^a$. 
The detail shows a small spacetime region within $R_a$ 
where the surface $\sigma$ advances through a spacetime cell at point $x$. Associated with the cell at point 
$x$ is the incremental spacetime volume $\rd \omega$ and the incremental Brownian variable $\rd \xi_x^a$.
}}
\end{center}
\end{figure}

We specify a probability space $(\Omega, {\cal F}, \mathbb{Q})$ along with a filtration 
${\cal F}^{\xi}_{\sigma}$ of ${\cal F}$ generated by a two-dimensional $\mathbb{Q}$-Brownian motion 
$\{\xi^{\it 1}_{\sigma},\xi^{\it 2}_{\sigma}\}$. For each interaction region $R_a$ ($a={\it 1},{\it 2}$) the spacelike hypersurfaces
$\{\sigma\}$ characterise the time evolution for each component of the Brownian motion. Given a foliation of spacetime, 
we define a ``time difference'' between any two surfaces as the spacetime volume enclosed by the surfaces 
within the region $R_a$. Consider the set $(\sigma_i, \sigma)$ of all spacetime points between the 
two spacelike surfaces $\sigma_i$ and $\sigma$, and consider the intersection of this set with the
interaction region $(\sigma_i,\sigma)\cap R_a$. We denote the spacetime volume of $(\sigma_i,\sigma)\cap R_a$
by $\omega_{\sigma}^{a}$ (see the gray shaded region in figure 2). The two volumes $\omega_{\sigma}^{\it 1}$ and $\omega_{\sigma}^{\it 2}$
correspond to two different time parameters for the two component Brownian motions. 
This definition ensures that time increases monotonically as the future surface ${\sigma}$ advances. The parameterization is covariant 
and has the convenience of only being relevant during the predefined measurement events. We define an 
infinitesimal increment of the Brownian motion $\rd \xi^a_x$ (relating to two spacelike hypersurfaces 
which differ only by an infinitesimal spacetime volume $\rd\omega$ at point $x$) by the following:
\bea
\rd \xi^a_x &=& 0, \;\; \text{for $x\notin R_a$}; \nonumber\\
\mathbb{E}^{\mathbb{Q}}[\rd \xi^a_x|{\cal F}_{\sigma}^{\xi}] &=& 0, \;\; \text{for $x$ to the future of $\sigma$}; \nonumber\\
\rd \xi^a_x \rd \xi^b_y &=& \delta^{ab}\delta_{xy}\rd\omega, \;\; \text{for $x\in R_a$, $y\in R_b$},
\label{BrownProps}
\eea
where $\mathbb{E}^{\mathbb{Q}}[\cdot|{\cal F}_{\sigma}^{\xi}]$ denotes conditional 
expectation in $\mathbb{Q}$. We attribute $\rd \xi^a_x$ to the spacetime point $x$
independent of any spacelike surface on which $x$ may lie. 
The two-dimensional Brownian motion is given by the sum of all infinitesimal Brownian increments
belonging to the set of points $(\sigma_i,\sigma)\cap R_a$,
\bea
\xi_{\sigma}^{a} = \int_{\sigma_i}^{\sigma} \rd \xi^a_x,
\label{BROWN}
\eea
so that an increment of the process can be written
\bea
\xi_{\sigma'}^{a}-\xi_{\sigma}^{a} = \int_{\sigma}^{\sigma'} \rd \xi^a_x,
\eea
where $\sigma'$ is to the future of $\sigma$.
These increments are independent and have mean zero and variance $\omega_{\sigma'}^{a}-\omega_{\sigma}^{a}$
as can easily be demonstrated by comparison with the conventional time parameterization of Brownian motion.

The state reduction process which occurs as the isospin state is measured can now be described by
extension of the Tomonaga equation (\ref{TOMdif}) to include a stochastic term. We define our evolution by 
\bea
\rd_x |\psi(\sigma)\rangle &=& \left\{ 2\lambda S_{\it 1} \rd \xi^{\it 1}_x  
-\half \lambda^2 \rd\omega  \right\}|\psi(\sigma)\rangle\;\; \text{for} \;\; x\in R_{\it 1},
\nonumber\\
\rd_x |\psi(\sigma)\rangle &=& \left\{ 2\lambda S_{\it 2} \rd \xi^{\it 2}_x 
-\half \lambda^2 \rd\omega  \right\}|\psi(\sigma)\rangle\;\; \text{for} \;\; x\in R_{\it 2},
\nonumber\\
\rd_x |\psi(\sigma)\rangle &=& 0 \;\; {\rm otherwise}.
\label{model}
\eea
The operators $S_a$ are isospin operators for each particle with the properties
\bea
S_{\it 1}|\pm\half; \;\cdot\; \rangle = \pm\half|\pm\half; \;\cdot\; \rangle \;\; , \;\;
S_{\it 2}|\;\cdot\; ; \pm\half \rangle = \pm\half|\; \cdot\; ; \pm\half \rangle;
\eea
the parameter $\lambda$ is a coupling parameter. The model explicitly describes an experiment to measure 
the isospin state of each particle in the given fixed isospin direction (the case of a general 
isospin measurement direction will be considered below). The form 
of equations~(\ref{model}) can be roughly understood by considering an incremental stage in the 
evolution where $\rd \xi^a_{\sigma}$ is either positive or negative. For example, if $\rd \xi^{\it 1}_{\sigma}$
is positive then the stochastic term on the right side of the first equation in (\ref{model}) will augment the 
$+\half$ state for particle ${\it 1}$ whilst degrading the $-\half$ state for particle ${\it 1}$.
The opposite happens if $\rd \xi^{\it 1}_{\sigma}$ is negative. Eventually after a certain period of evolution 
one of the two eigenstates will dominate. This is analogous to the famous problem of the gambler's ruin. 

The drift terms on the right side of equations (\ref{model}) ensure that the state norm is a positive martingale
\bea
\rd_x \langle \psi (\sigma) |\psi(\sigma)\rangle &=& 4\lambda 
\langle \psi (\sigma)|S_{\it 1}|\psi(\sigma)\rangle \rd\xi_x^{\it 1}  \;\; \text{for} \;\; x\in R_{\it 1},
\nonumber\\
\rd_x \langle \psi (\sigma) |\psi(\sigma)\rangle &=& 4\lambda 
\langle \psi (\sigma)|S_{\it 2}|\psi(\sigma)\rangle \rd\xi_x^{\it 2}  \;\; \text{for} \;\; x\in R_{\it 2}.
\eea
We can then define a physical measure $\mathbb{P}$ equivalent to $\mathbb{Q}$ according to
\bea
\mathbb{E}^{\mathbb{P}} [\cdot | {\cal F}_{\sigma}^{\xi}] = 
\frac{\mathbb{E}^{\mathbb{Q}} [\langle \psi (\sigma_f) |\psi(\sigma_f)\rangle \cdot | {\cal F}_{\sigma}^{\xi}] }
{\mathbb{E}^{\mathbb{Q}} [\langle \psi(\sigma_f) |\psi(\sigma_f)\rangle | {\cal F}_{\sigma}^{\xi}] }
=\frac{\mathbb{E}^{\mathbb{Q}} [\langle \psi (\sigma_f) |\psi(\sigma_f)\rangle \cdot | {\cal F}_{\sigma}^{\xi}]}
{\langle \psi(\sigma) |\psi(\sigma)\rangle},
\label{measchange}
\eea
with $\sigma_f$ the final surface of the state evolution we are considering.
This change of measure ensures that physical outcomes are weighted according to the Born rule, meeting
the second bullet-pointed criterion for dynamical state reduction stated in the introduction. Note that
the processes $\xi_{\sigma}^a$ satisfy a modified distribution under the $\mathbb{P}$-measure.

Our model can be interpreted as an effective model describing the interaction
of the two particles with macroscopic measuring devices in regions
$R_{\it 1}$ and $R_{\it 2}$. In more detail we would expect the particle states to become correlated
with different states of the measuring devices. The state reduction dynamics would be expected to 
have a negligible effect on the individual spin particles, however, the effect would be rapid for a 
macroscopic superposition of measuring device states. Collapse of the spin particle 
would then occur indirectly as a result of collapse of the macro state. In our model we have assumed that the particle 
states undergo a direct collapse dynamics. This allows us to ignore the fine details of the interaction 
between spin particles and measuring devices. 

By designating spacetime regions where collapse of the isospin state occurs we avoid the issue 
of setting a scale distinguishing micro and macro behaviour. Our main interest here is to
understand the dynamical process of state reduction for an entangled quantum system in a 
relativistic setting.

\section{Solution in terms of $\mathbb{Q}$-Brownian motion}
\label{sec:qsol}

Working in the $\mathbb{Q}$-measure where $\xi_{\sigma}^a$ is a Brownian process we find the following solution 
for the unnormalised state evolution:
\bea
|\psi(\sigma)\rangle=\sqrtwo\left\{
e^{\lambda \xi_{\sigma}^{\it 1}-\lambda^2\omega_{\sigma}^{\it 1}}
e^{-\lambda \xi_{\sigma}^{\it 2}-\lambda^2\omega_{\sigma}^{\it 2}}|+\half;-\half\rangle
-e^{-\lambda \xi_{\sigma}^{\it 1}-\lambda^2\omega_{\sigma}^{\it 1}}
e^{\lambda \xi_{\sigma}^{\it 2}-\lambda^2\omega_{\sigma}^{\it 2}}|-\half;+\half\rangle
\right\}.
\label{qsol}
\eea
This can easily be checked with the use of (\ref{BrownProps}), (\ref{BROWN}), and (\ref{model}).
The state norm is given by
\bea
\langle\psi(\sigma)|\psi(\sigma)\rangle=\half\left\{
e^{2\lambda \xi_{\sigma}^{\it 1}-2\lambda^2\omega_{\sigma}^{\it 1}}
e^{-2\lambda \xi_{\sigma}^{\it 2}-2\lambda^2\omega_{\sigma}^{\it 2}}
+e^{-2\lambda \xi_{\sigma}^{\it 1}-2\lambda^2\omega_{\sigma}^{\it 1}}
e^{2\lambda \xi_{\sigma}^{\it 2}-2\lambda^2\omega_{\sigma}^{\it 2}}
\right\}.
\label{normq}
\eea
We note that although equation (\ref{qsol}) is a solution to (\ref{model}), it cannot be considered as a 
solution to the model since it completely disregards the important role played by the physical measure~$\mathbb{P}$. 
Equation (\ref{qsol}) enables us to generate sample outcomes, however, the physical probability 
density at a given outcome can only be determined afterwards with reference to the state norm (a likely 
outcome in $\mathbb{Q}$ may be highly unlikely in $\mathbb{P}$).   

We define the characteristic function associated with $\xi_{\sigma}^{\it 1}$ and $\xi_{\sigma}^{\it 2}$
in the $\mathbb{P}$-measure as
\bea
\Phi_{\sigma}^{\xi}(t_{\it 1},t_{\it 2}) &=& \mathbb{E}^{\mathbb{P}} \left[ e^{it_{\it 1}\xi_{\sigma}^{\it 1}} 
e^{it_{\it 2}\xi_{\sigma}^{\it 2}} | {\cal F}_{\sigma_i}^{\xi} \right]
\label{pchar}\\
&=& \mathbb{E}^{\mathbb{Q}} \left[ \langle \psi(\sigma) |\psi(\sigma)\rangle e^{it_{\it 1}\xi_{\sigma}^{\it 1}} 
e^{it_{\it 2}\xi_{\sigma}^{\it 2}}| {\cal F}_{\sigma_i}^{\xi} \right],
\eea
where we have used equation (\ref{measchange}) and the fact that the initial state has unit norm. 
Noting that $\xi_{\sigma}^{\it 1}$ and $\xi_{\sigma}^{\it 2}$ are independent in the $\mathbb{Q}$-measure we can determine 
the expectation using equation (\ref{normq}) to find
\bea
\Phi_{\sigma}^{\xi}(t_{\it 1},t_2)=\half\left\{
e^{2i\lambda t_{\it 1} \omega_{\sigma}^{\it 1}- \frac{1}{2} t_{\it 1}^2\omega_{\sigma}^{\it 1}}
e^{-2i\lambda t_{\it 2} \omega_{\sigma}^{\it 2}- \frac{1}{2} t_{\it 2}^2\omega_{\sigma}^{\it 2}}
+e^{-2i\lambda t_{\it 1} \omega_{\sigma}^{\it 1}- \frac{1}{2} t_{\it 1}^2\omega_{\sigma}^{\it 1}}
e^{2i\lambda t_{\it 2} \omega_{\sigma}^{\it 2}- \frac{1}{2} t_{\it 2}^2\omega_{\sigma}^{\it 2}}
\right\}.
\label{charfun}
\eea 
The characteristic function allows us to immediately demonstrate that spacelike separated processes
$\xi_{\sigma}^{1}$ and $\xi_{\sigma}^{2}$ are correlated under the physical measure $\mathbb{P}$:
\bea
\mathbb{E}^{\mathbb{P}}\left[ \xi_{\sigma}^{a} | {\cal F}_{\sigma_i}^{\xi} \right] &=& -i\frac{\rd}{\rd t_a}\left.\left[
\Phi_{\sigma}^{\xi}(t_{\it 1},t_{\it 2})\right]\right|_{t_{\it 1} = t_{\it 2} = 0} = 0,\nonumber\\
\mathbb{E}^{\mathbb{P}}\left[ \xi_{\sigma}^{\it 1} \xi_{\sigma}^{\it 2}| {\cal F}_{\sigma_i}^{\xi} \right] 
&=& -\frac{\rd^2}{\rd t_{\it 1} \rd t_{\it 2}}\left.\left[
\Phi_{\sigma}^{\xi}(t_{\it 1},t_{\it 2})\right]\right|_{t_{\it 1} = t_{\it 2} = 0} 
= -4\lambda^2 \omega^{\it 1}_{\sigma}\omega^{\it 2}_{\sigma}.
\eea
The stochastic information at one wing of the apparatus is not independent of the stochastic information at 
the other wing. We might expect this since the results of the two measurements that the information dictate 
are correlated.

Before demonstrating the state reducing properties of this model, we first show in the next section how to 
express the solution (\ref{qsol}) directly in terms of a $\mathbb{P}$-Brownian motion. This will allow us to 
generate physical sample solutions.

\section{Solution in terms of $\mathbb{P}$-Brownian motion}
\label{sec:psol}

Let the probability space $(\Omega, {\cal F}, \mathbb{P})$ be given and let ${\cal G}_{\sigma}$ be a 
filtration of ${\cal F}$ such that independent $\mathbb{P}$-Brownian motions $B_{\sigma}^a$ ($a={\it 1},{\it 2}$) 
are specified together with random variables $s_{a}$ (independent of $B_{\sigma}^a$). The Brownian motions
$B_{\sigma}^a$ are defined under the $\mathbb{P}$-measure in the same way in which Brownian motions 
$\xi_{\sigma}^a$ are defined under $\mathbb{Q}$-measure by equations~(\ref{BrownProps}) and~(\ref{BROWN}).
The probability distribution for the random variables $s_{a}$ are given by
\bea
\mathbb{P}\left( s_{\it 1} = +\half , s_{\it 2} = -\half \right) &=& \half, \nonumber\\
\mathbb{P}\left( s_{\it 1} = -\half , s_{\it 2} = +\half \right) &=& \half. 
\label{probs}
\eea
We assume that $s_{a}$ are ${\cal G}_{\sigma_i}$-measurable. 

Now define the random processes (c.f. \cite{Dorj2})
\bea
\xi_{\sigma}^{\it 1}&=&4\lambda s_{\it 1}\omega_{\sigma}^{\it 1} + B_{\sigma}^{\it 1},\nonumber\\
\xi_{\sigma}^{\it 2}&=&4\lambda s_{\it 2}\omega_{\sigma}^{\it 2} + B_{\sigma}^{\it 2}.
\label{xiproc}
\eea
Our aim is to show that these processes, defined under the $\mathbb{P}$-measure, can be identified as the 
$\mathbb{Q}$-Brownian processes $\xi_{\sigma}^{a}$ involved in the equations of motion for the state (\ref{model}). In order to do
this we must show that their characteristic function under the $\mathbb{P}$-measure is identical to that found 
for the $\mathbb{Q}$-Brownian processes, as given by equation (\ref{charfun}).

Again let ${\cal F}_{\sigma}^{\xi}$ denote the filtration generated by $\{\xi_{\sigma}^{\it 1},\xi_{\sigma}^{\it 2}\}$. 
The use of ${\cal F}_{\sigma}^{\xi}$ ensures that we have no more or less information than is given by the processes 
$\{\xi_{\sigma}^{\it 1}, \xi_{\sigma}^{\it 2}\}$ as in the original presentation of the model in section \ref{sec:model}. Neither $s_a$ 
nor $B_{\sigma}^a$ are ${\cal F}_{\sigma}^{\xi}$-measurable. The only information we have regarding the 
realisation of these variables is $\{\xi_{\sigma}^{\it 1},\xi_{\sigma}^{\it 2}\}$.

The characteristic function for $\xi_{\sigma}^{\it 1}$ and $\xi_{\sigma}^{\it 2}$ is given by equation (\ref{pchar}),
\bea
\Phi_{\sigma}^{\xi}(t_{\it 1},t_{\it 2}) = \mathbb{E}^{\mathbb{P}} \left[ e^{it_{\it 1}\xi_{\sigma}^{\it 1}}
 e^{it_{\it 2}\xi_{\sigma}^{\it 2}} | {\cal F}_{\sigma_i}^{\xi}\right],
\nonumber
\eea
but now we write
\bea
\Phi_{\sigma}^{\xi}(t_{\it 1},t_{\it 2}) &=& 
\half\mathbb{E}^{\mathbb{P}} \left[\left. e^{it_{\it 1}(4\lambda s_{\it 1}\omega_{\sigma}^{\it 1} + B_{\sigma}^{\it 1})} 
e^{it_{\it 2}(4\lambda s_{\it 2}\omega_{\sigma}^{\it 2} + B_{\sigma}^{\it 2})}\right| 
{\cal F}_{\sigma_i}^{\xi}; s_{\it 1}=+\half , s_{\it 2} = -\half \right]
\nonumber\\
&&+\half\mathbb{E}^{\mathbb{P}} \left[\left. e^{it_{\it 1}(4\lambda s_{\it 1}\omega_{\sigma}^{\it 1} + B_{\sigma}^{\it 1})} 
e^{it_{\it 2}(4\lambda s_{\it 2}\omega_{\sigma}^{\it 2} + B_{\sigma}^{\it 2})}\right| 
{\cal F}_{\sigma_i}^{\xi}; s_{\it 1}=-\half , s_{\it 2} = +\half \right].
\eea
Noting that $B_{\sigma}^{\it 1}$ and $B_{\sigma}^{\it 2}$ are independent we can work directly in the $\mathbb{P}$-measure
to confirm that the characteristic function is once more given by equation (\ref{charfun}). This demonstrates that the processes 
defined by equation (\ref{xiproc}) can indeed be identified as $\mathbb{Q}$-Brownian motions $\xi_{\sigma}^{a}$.

We are now in a position to express the solution to equations (\ref{model}) and (\ref{measchange}) in terms
of the $\mathbb{P}$-Brownian motions $B_{\sigma}^{a}$, and the random variables $s_a$. This 
is summarised in the following subsection. The fact that the solution is expressed in terms of variables with an 
{\it a priori} known probability distribution in the physical measure is to be contrasted with the solution in terms
of $\mathbb{Q}$-Brownian motion where physical probabilities can only be determined {\it a posteriori} with 
knowledge of the state norm.
 
\subsection{Summary of solution}
The solution to the equations of motion (\ref{model}) is given by the unnormalised state
\bea
|\psi(\sigma)\rangle=\sqrtwo\left\{
e^{\lambda \xi_{\sigma}^{\it 1}-\lambda^2\omega_{\sigma}^{\it 1}}
e^{-\lambda \xi_{\sigma}^{\it 2}-\lambda^2\omega_{\sigma}^{\it 2}}|+\half;-\half\rangle
-e^{-\lambda \xi_{\sigma}^{\it 1}-\lambda^2\omega_{\sigma}^{\it 1}}
e^{\lambda \xi_{\sigma}^{\it 2}-\lambda^2\omega_{\sigma}^{\it 2}}|-\half;+\half\rangle
\right\}.
\label{soln}
\eea
(This is the same solution in terms of $\xi_{\sigma}^{a}$ as presented in equation (\ref{qsol}), however, we now
treat $\xi_{\sigma}^{a}$, not as a $\mathbb{Q}$-Brownian motion, but as an information process defined in terms
of variables with known $\mathbb{P}$-distributions.) The random variables $\xi_{\sigma}^{a}$ are given by
\bea
\xi_{\sigma}^{\it 1}&=&4\lambda s_{\it 1}\omega_{\sigma}^{\it 1} + B_{\sigma}^{\it 1},\nonumber\\
\xi_{\sigma}^{\it 2}&=&4\lambda s_{\it 2}\omega_{\sigma}^{\it 2} + B_{\sigma}^{\it 2}.
\eea
The stochastic processes $B_{\sigma}^{\it 1}$ and $B_{\sigma}^{\it 2}$ are independent $\mathbb{P}$-Brownian motions. The random variables 
$s_{a}$ take values $s_{\it 1} =+1/2, s_{\it 2} =-1/2$ with probability $1/2$ and 
$s_{\it 1} =-1/2, s_{\it 2} =+1/2$ with probability $1/2$. Brownian motions $B_{\sigma}^a$ and random variables $s^{a}$ are 
independent. Only the processes $\xi_{\sigma}^{a}$ are measurable.

This solution is as relativistically invariant as a description of state reduction can be. We expect the state to 
depend on the spacelike surface $\sigma$ we choose to query. The dependence on $\sigma$ results in equation (\ref{soln}) from the 
spacetime volume variables $\omega_{\sigma}^a$ and the random variables $B_{\sigma}^a$. We note that neither 
of these variables depends on the chosen foliation of spacetime. For example, the distribution 
of $B_{\sigma}^a$ is characterized by the spacetime volume $\omega_{\sigma}^a$ which in turn is determined 
only by the surface $\sigma$. A foliation dependence would be undesirable as it would 
indicate a preferred frame in the model. The fact that there is no foliation dependence indicates also that the 
choice $\sigma$ has no prior physical significance.

\subsection{State reduction}
\label{SR}
In this subsection we explicitly demonstrate how the solution outlined above exhibits state 
reduction to a state of well-defined isospin. Consider the isospin operators $S_a$. The 
conditional expectation of $S_a$ for the state $|\psi(\sigma)\rangle$ is given by
\bea
\langle S_{a}\rangle_{\sigma}  = \frac{\langle \psi(\sigma)|S_a |\psi(\sigma)\rangle}
{\langle \psi(\sigma) | \psi(\sigma)\rangle}.
\eea
From equation (\ref{soln}) we find choosing, for example, $a={\it 1}$,
\bea
\langle S_{\it 1}\rangle_{\sigma}=
\frac{ \half e^{2\lambda \xi_{\sigma}^{\it 1}-2\lambda^2\omega_{\sigma}^{\it 1}} e^{-2\lambda \xi_{\sigma}^{\it 2}-2\lambda^2\omega_{\sigma}^{\it 2}}
-\half e^{-2\lambda \xi_{\sigma}^{\it 1}-2\lambda^2\omega_{\sigma}^{\it 1}} e^{2\lambda \xi_{\sigma}^{\it 2}-2\lambda^2\omega_{\sigma}^{\it 2}}}
{ e^{2\lambda \xi_{\sigma}^{\it 1}-2\lambda^2\omega_{\sigma}^{\it 1}} e^{-2\lambda \xi_{\sigma}^{\it 2}-2\lambda^2\omega_{\sigma}^{\it 2}}
+e^{-2\lambda \xi_{\sigma}^{\it 1}-2\lambda^2\omega_{\sigma}^{\it 1}} e^{2\lambda \xi_{\sigma}^{\it 2}-2\lambda^2\omega_{\sigma}^{\it 2}}}.
\label{Iexp}
\eea
Now suppose we condition on the event $s_{\it 1} =+1/2, s_{\it 2} =-1/2$. We find
\bea
\langle S_{\it 1}\rangle_{\sigma}&=&
\frac{ \half e^{2\lambda B_{\sigma}^{\it 1}+2\lambda^2\omega_{\sigma}^{\it 1}} e^{-2\lambda B_{\sigma}^{\it 2}+2\lambda^2\omega_{\sigma}^{\it 2}}
-\half e^{-2\lambda B_{\sigma}^{\it 1}-6\lambda^2\omega_{\sigma}^{\it 1}} e^{2\lambda B_{\sigma}^{\it 2}-6\lambda^2\omega_{\sigma}^{\it 2}}}
{ e^{2\lambda B_{\sigma}^{\it 1}+2\lambda^2\omega_{\sigma}^{\it 1}} e^{-2\lambda B_{\sigma}^{\it 2}+2\lambda^2\omega_{\sigma}^{\it 2}}
+e^{-2\lambda B_{\sigma}^{\it 1}-6\lambda^2\omega_{\sigma}^{\it 1}} e^{2\lambda B_{\sigma}^{\it 2}-6\lambda^2\omega_{\sigma}^{\it 2}}}
\nonumber\\
&=& \frac{ \half
-\half e^{-4\lambda B_{\sigma}^{\it 1}-8\lambda^2\omega_{\sigma}^{\it 1}} e^{4\lambda B_{\sigma}^{\it 2}-8\lambda^2\omega_{\sigma}^{\it 2}}}
{ 1+e^{-4\lambda B_{\sigma}^{\it 1}-8\lambda^2\omega_{\sigma}^{\it 1}} e^{4\lambda B_{\sigma}^{\it 2}-8\lambda^2\omega_{\sigma}^{\it 2}}}.
\eea
Next we use the result that
\bea
\lim_{\omega_{\sigma}\rightarrow\infty} \mathbb{P}\left(
e^{\pm4\lambda B_{\sigma}-8\lambda^2\omega_{\sigma}} > 0
\right) = 0,
\label{problimit}
\eea
to deduce that $\langle S_{\it 1}\rangle_{\sigma} \rightarrow 1/2$ as $\omega_{\sigma}^{\it 1} \rightarrow \infty$
or $\omega_{\sigma}^{\it 2} \rightarrow \infty$. These volumes increase in size as the surface $\sigma$ 
passes the spacetime regions $R_{\it 1}$ and $R_{\it 2}$ respectively. Since these regions are of finite size, $\omega_{\sigma}^{\it 1}$ and 
$\omega_{\sigma}^{\it 1}$ can only attain fixed maximal values. We assume that these maximal values are sufficiently large
that the limit of equation (\ref{problimit}) is approached with high precision.
Note that the rate at which this limit is approached can be controlled by the choice of coupling parameter $\lambda$. 

A similar analysis leads to the conclusion that $\langle S_{\it 2}\rangle_{\sigma} \rightarrow -1/2$. Conversely, 
if we were to condition on the event $s_{\it 1} =-1/2, s_{\it 2} =+1/2$, we would find 
$\langle S_{\it 1}\rangle_{\sigma} \rightarrow -1/2$ and $\langle S_{\it 2}\rangle_{\sigma} \rightarrow 1/2$. 
We observe that the unmeasurable random variable $s_a$ dictates the outcome of the experiment. Only the processes 
$\xi_{\sigma}^a$ are known to the state; the Brownian processes $B_{\sigma}^a$ act as noise terms obscuring the 
values $s_a$.

\subsection{Probabilities for reduction}
Here we demonstrate that the stochastic probabilities for outcomes are those predicted by the
quantum state prior to the measurement event. For example, we define the $+\half$ state projection 
operator on particle ${\it 1}$ by 
\bea
P_{\it 1}^+|+\half; \;\cdot\; \rangle = |+\half; \;\cdot\; \rangle \;\; ; \;\;
P_{\it 1}^+|-\half; \;\cdot\; \rangle = 0,
\eea
and the conditional expectation of this operator for the state $|\psi(\sigma)\rangle$ by
\bea
\langle P_{\it 1}^+\rangle_{\sigma}  = \frac{\langle \psi(\sigma)|P_{\it 1}^+ |\psi(\sigma)\rangle}
{\langle \psi(\sigma) | \psi(\sigma)\rangle}.
\eea
In order to calculate the unconditional expectation of $\langle P_{\it 1}^+\rangle_{\sigma}$ it turns out to be simpler to work in the 
$\mathbb{Q}$-measure. We proceed as follows:
\bea
\mathbb{E}^{\mathbb{P}} [\langle P_{\it 1}^+\rangle_{\sigma}| {\cal F}_{\sigma_i}^{\xi}] &=& 
\mathbb{E}^{\mathbb{Q}} [\langle \psi(\sigma) |\psi(\sigma)\rangle \langle P_{\it 1}^+\rangle_{\sigma}
| {\cal F}_{\sigma_i}^{\xi}] 
\nonumber\\
&=&\mathbb{E}^{\mathbb{Q}} [\langle \psi (\sigma)|P_{\it 1}^+ |\psi(\sigma)\rangle| {\cal F}_{\sigma_i}^{\xi}] 
\nonumber\\
&=&\mathbb{E}^{\mathbb{Q}} \left.\left[\half
e^{2\lambda \xi_{\sigma}^{\it 1}-2\lambda^2\omega_{\sigma}^{\it 1}}
e^{-2\lambda \xi_{\sigma}^{\it 2}-2\lambda^2\omega_{\sigma}^{\it 2}} \right| {\cal F}_{\sigma_i}^{\xi}\right] 
=\half.
\label{proexp}
\eea
From the previous subsection we know that as $\omega_{\sigma}^{a}\rightarrow\infty$ then the state of each particle
tends towards a definite isospin state and consequently the conditional expectation of $P_{\it 1}^+$ tends to 
either 0 or 1. This means that as $\omega_{\sigma}^{a}\rightarrow\infty$ we have 
\bea
\mathbb{E}^{\mathbb{P}} [\langle P_{\it 1}^+\rangle_{\sigma}| {\cal F}_{\sigma_i}^{\xi}] 
=\mathbb{E}^{\mathbb{P}} \left.\left[\mathds{1}_{\left\{\langle S_{\it 1}\rangle_{\sigma}=\half\right\}}
\right| {\cal F}_{\sigma_i}^{\xi}\right] 
=\mathbb{P}\left.\left(\langle S_{\it 1}\rangle_{\sigma}=\half \right| {\cal F}_{\sigma_i}^{\xi}\right), 
\eea
where $\mathds{1}_{\{E\}}$ takes the value 1 if the event $E$ is true, and 0 otherwise. From equation (\ref{proexp}) we 
can now write
\bea
\mathbb{P}\left.\left(\langle S_{\it 1}\rangle_{\sigma}=\half \right| {\cal F}_{\sigma_i}^{\xi}\right)
=\half =\langle P_{\it 1}^+\rangle_{\sigma_i}.
\eea
This tells us that as the dynamics lead to a definite state for each particle then the stochastic 
probability of a given outcome matches the initial quantum probability. The same is true of other 
projection operators as can easily be shown.

\section{Interpretation in terms of nonlinear filtering}
\label{sec:filter}

In this section we use the method of Brody and Hughston \cite{Dorj, Dorj2} to demonstrate that the problem under 
consideration can be interpreted as a classical nonlinear filtering problem. The method was originally
applied to solve an energy-based state diffusion equation.

From section \ref{SR} we understand that the ${\cal F}_{\sigma}^{\xi}$-unmeasurable random variables $s_a$ 
represent the true outcomes for the isospin eigenvalues of each particle after the measurement process.
Only information in the form $\xi_{\sigma}^a=4\lambda s_{a}\omega_{\sigma}^a + B_{\sigma}^{a}$ is 
accessible to the state where the realised value of $s_a$ is masked by the ${\cal F}_{\sigma}^{\xi}$-unmeasurable 
noise processes $B_{\sigma}^{a}$.

Suppose we attempt to address the problem of finding $s_a$ directly, that is, given 
$\{\xi_{\sigma}^a\}$ what is the best estimate we can make for $s_a$. This is a classical nonlinear 
filtering problem. It is straightforward to show that the best estimate for the value of $s_{a}$ is 
given by the conditional expectation
\bea
\widehat{s_{a}}_{\sigma} = \mathbb{E}^{\mathbb{P}}\left[\left. s_{a} \right|{\cal F}_{\sigma}^{\xi}\right].
\label{filter}
\eea
The aim is now to identify $\widehat{s_{a}}_{\sigma}$ with the quantum expectation processes
$\langle S_{a} \rangle_{\sigma}$.

We first show that $\xi_{\sigma}^{a}$ are Markov processes. To do this we show that 
\bea
\mathbb{P}\left(\left. \xi_{\sigma}^{a} < y \right| 
\xi_{\sigma_1}^{\it 1},\xi_{\sigma_2}^{\it 1},\cdots, \xi_{\sigma_k}^{\it 1} ; 
\xi_{\sigma_1}^{\it 2},\xi_{\sigma_2}^{\it 2},\cdots, \xi_{\sigma_k}^{\it 2}
\right) =
\mathbb{P}\left(\left. \xi_{\sigma}^{a} < y \right| \xi_{\sigma_1}^{\it 1} ; \xi_{\sigma_1}^{\it 2} \right)
\label{markov}
\eea
where $\{\sigma,\sigma_1,\sigma_2,\cdots,\sigma_k\}$ is a sequence of spacelike surfaces belonging to some 
spacetime foliation such that
\bea
\omega_{\sigma}^{\it 1} &\geq& \omega_{\sigma_1}^{\it 1} \geq \omega_{\sigma_2}^{\it 1} \geq \cdots \geq \omega_{\sigma_k}^{\it 1}> 0,
\nonumber\\
\omega_{\sigma}^{\it 2} &\geq& \omega_{\sigma_1}^{\it 2} \geq \omega_{\sigma_2}^{\it 2} \geq \cdots \geq \omega_{\sigma_k}^{\it 2}> 0.
\eea
The proof of equation (\ref{markov}) is more or less identical to that given by Brody and Hughston~\cite{Dorj}.
We use the fact that $\mathbb{E}^{\mathbb{P}}[B_{\sigma'}^{b}B_{\sigma''}^{b}]= \omega_{\sigma'}^{b}$, where 
$\omega_{\sigma''}^{b} \geq \omega_{\sigma'}^{b}$ for $b={\it 1},{\it 2}$. Then for 
$\omega_{\sigma}^{b} \geq \omega_{\sigma_1}^{b} \geq \omega_{\sigma_2}^{b} > 0$
we have that
\bea
B_{\sigma}^{b} {\rm  \;\; and \;\;} \frac{B_{\sigma_1}^{b}}{\omega_{\sigma_1}^{b}} - \frac{B_{\sigma_2}^{b}}{\omega_{\sigma_2}^{b}}
{\rm \;\; are \;\; independent}.
\label{indep1}
\eea
Furthermore,
\bea
\frac{B_{\sigma_1}^{b}}{\omega_{\sigma_1}^{b}} - \frac{B_{\sigma_2}^{b}}{\omega_{\sigma_2}^{b}}
= \frac{\xi_{\sigma_1}^{b}}{\omega_{\sigma_1}^{b}} - \frac{\xi_{\sigma_2}^{b}}{\omega_{\sigma_2}^{b}},
\eea
from which it follows that 
\bea
&&\mathbb{P}\left(\left. \xi_{\sigma}^{a} < y \right| \xi_{\sigma_1}^{\it 1},\xi_{\sigma_2}^{\it 1}, \xi_{\sigma_3}^{\it 1}, \cdots
; \xi_{\sigma_1}^{\it 2},\xi_{\sigma_2}^{\it 2},\xi_{\sigma_3}^{\it 2},\cdots\right) \nonumber\\
&=& \mathbb{P}\left(\left. \xi_{\sigma}^{a} < y \right| \xi_{\sigma_1}^{\it 1} , 
\frac{\xi_{\sigma_1}^{\it 1}}{\omega_{\sigma_1}^{\it 1}} - \frac{\xi_{\sigma_2}^{\it 1}}{\omega_{\sigma_2}^{\it 1}}, 
\frac{\xi_{\sigma_2}^{\it 1}}{\omega_{\sigma_2}^{\it 1}} - \frac{\xi_{\sigma_3}^{\it 1}}{\omega_{\sigma_3}^{\it 1}},\cdots;
\xi_{\sigma_1}^{\it 2} , 
\frac{\xi_{\sigma_1}^{\it 2}}{\omega_{\sigma_1}^{\it 2}} - \frac{\xi_{\sigma_2}^{\it 2}}{\omega_{\sigma_2}^{\it 2}}, 
\frac{\xi_{\sigma_2}^{\it 2}}{\omega_{\sigma_2}^{\it 2}} - \frac{\xi_{\sigma_3}^{\it 2}}{\omega_{\sigma_3}^{\it 2}},\cdots
\right)\nonumber\\
&=& \mathbb{P}\left(\left. \xi_{\sigma}^{a} < y \right| \xi_{\sigma_1}^{\it 1} , 
\frac{B_{\sigma_1}^{\it 1}}{\omega_{\sigma_1}^{\it 1}} - \frac{B_{\sigma_2}^{\it 1}}{\omega_{\sigma_2}^{\it 1}}, 
\frac{B_{\sigma_2}^{\it 1}}{\omega_{\sigma_2}^{\it 1}} - \frac{B_{\sigma_3}^{\it 1}}{\omega_{\sigma_3}^{\it 1}},\cdots;
\xi_{\sigma_1}^{\it 2} , 
\frac{B_{\sigma_1}^{\it 2}}{\omega_{\sigma_1}^{\it 2}} - \frac{B_{\sigma_2}^{\it 2}}{\omega_{\sigma_2}^{\it 2}}, 
\frac{B_{\sigma_2}^{\it 2}}{\omega_{\sigma_2}^{\it 2}} - \frac{B_{\sigma_3}^{\it 2}}{\omega_{\sigma_3}^{\it 2}},\cdots
\right).
\eea
Now from (\ref{indep1}) we have that $\xi_{\sigma}^{a}$, $\xi_{\sigma_1}^{\it 1}$, and $\xi_{\sigma_1}^{\it 2}$ are each independent of 
$B_{\sigma_1}^{\it 1}/\omega_{\sigma_1}^{\it 1} - B_{\sigma_2}^{\it 1}/\omega_{\sigma_2}^{\it 1}$, $B_{\sigma_2}^{\it 1}/\omega_{\sigma_2}^{\it 1} - B_{\sigma_3}^{\it 1}/\omega_{\sigma_3}^{\it 1}$, etc. Equation (\ref{markov}) follows.
The same argument shows that 
\bea
\mathbb{P}\left(\left. B_{\sigma}^{a} < y \right| 
\xi_{\sigma_1}^{\it 1},\xi_{\sigma_2}^{\it 1},\cdots, \xi_{\sigma_k}^{\it 1} ; 
\xi_{\sigma_1}^{\it 2},\xi_{\sigma_2}^{\it 2},\cdots, \xi_{\sigma_k}^{\it 2}
\right) =
\mathbb{P}\left(\left. B_{\sigma}^{a} < y \right| \xi_{\sigma_1}^{\it 1} ; \xi_{\sigma_1}^{\it 2} \right),
\eea
and therefore
\bea
\mathbb{P}\left(\left. s_a = \pm\half \right| {\cal F}_{\sigma}^{\xi}\right) =
\mathbb{P}\left(\left. s_a = \pm\half \right| \xi_{\sigma}^{\it 1} ; \xi_{\sigma}^{\it 2} \right).
\eea
Next we use a version of Bayes formula to calculate this conditional probability
\bea
\mathbb{P}\left(\left. s_{\it 1} = \pm\half, s_{\it 2} = \mp\half \right| \xi_{\sigma}^{\it 1} ; \xi_{\sigma}^{\it 2} \right) = 
\frac{\mathbb{P}\left( s_{\it 1} = \pm\half, s_{\it 2} =\mp\half \right) 
\rho\left(\left. \xi_{\sigma}^{\it 1} ; \xi_{\sigma}^{\it 2}\right| s_{\it 1} = \pm\half, s_{\it 2}=\mp\half \right)}
{\rho\left(\xi_{\sigma}^{\it 1} ; \xi_{\sigma}^{\it 2}\right)}.
\eea
The density function for the random variables $(\xi_{\sigma}^{\it 1} ; \xi_{\sigma}^{\it 2})$ conditional on $s_a$ is
Gaussian (since $B_{\sigma}^a$ is a Brownian motion under $\mathbb{P}$) and is
given by
\bea
\rho\left(\left. \xi_{\sigma}^{\it 1} ; \xi_{\sigma}^{\it 2} \right| s_{\it 1} = \pm\half, s_{\it 2}=\mp\half \right) &\propto& 
e^{-\frac{1}{2\omega_{\sigma}^{\it 1} } \left( \xi_{\sigma}^{\it 1} \mp 2\lambda \omega_{\sigma}^{\it 1} \right)^2 }
e^{-\frac{1}{2\omega_{\sigma}^{\it 2} } \left( \xi_{\sigma}^{\it 2} \pm 2\lambda \omega_{\sigma}^{\it 2} \right)^2 }.
\eea
We also have that 
\bea
\rho\left(\xi_{\sigma}^{\it 1} ; \xi_{\sigma}^{\it 2} \right) = 
\half \rho\left(\left. \xi_{\sigma}^{\it 1} , \xi_{\sigma}^{\it 2}\right| s_{\it 1} = +\half, s_{\it 2}=-\half\right)
+\half \rho\left(\left. \xi_{\sigma}^{\it 1} , \xi_{\sigma}^{\it 2}\right| s_{\it 1} = -\half, s_{\it 2}=+\half\right).
\eea
We are now in a position to calculate the conditional expectation $\widehat{s_{a}}_{\sigma}$ given by equation (\ref{filter}).
For example, choosing $a={\it 1}$ we have
\bea
\widehat{s_{\it 1}}_{\sigma}  = \mathbb{E}^{\mathbb{P}}\left[\left. s_{\it 1} \right|{\cal F}_{\sigma}^{\xi}\right]
&=& \half \mathbb{P}\left(\left. s_{\it 1} = +\half, s_{\it 2} = -\half \right| \xi_{\sigma}^{\it 1} ; \xi_{\sigma}^{\it 2} \right) 
- \half \mathbb{P} \left(\left. s_{\it 1} = -\half, s_{\it 2} = +\half \right| \xi_{\sigma}^{\it 1} ; \xi_{\sigma}^{\it 2} \right) 
\nonumber\\
&=&
\frac{
\half e^{2\lambda \xi_{\sigma}^{\it 1}}e^{-2\lambda \xi_{\sigma}^{\it 2}}
-\half e^{-2\lambda \xi_{\sigma}^{\it 1}}e^{2\lambda \xi_{\sigma}^{\it 2}}}
{
e^{2\lambda \xi_{\sigma}^{\it 1}}e^{-2\lambda \xi_{\sigma}^{\it 2}}
+e^{-2\lambda \xi_{\sigma}^{\it 1}}e^{2\lambda \xi_{\sigma}^{\it 2}}}.
\eea
This is the same expression as that given for $\langle S_{\it 1} \rangle_{\sigma}$ in 
equation (\ref{Iexp}). This demonstrates that the conditional expectation $\widehat{s_{\it 1}}_{\sigma}$, which 
represents our best estimate for the random variable $s_{\it 1}$ given only information from the filtration
${\cal F}^\xi_{\sigma}$, corresponds to the quantum expectation of the operator $S_{\it 1}$, conditional
on the same information. It is remarkable that the complexity of the stochastic quantum formalism
corresponds to a such a conceptually intuitive classical analogue.

\section{Bell test experiments}
\label{sec:bell}

\begin{figure}[t]
\begin{center}
\includegraphics[width=8cm]{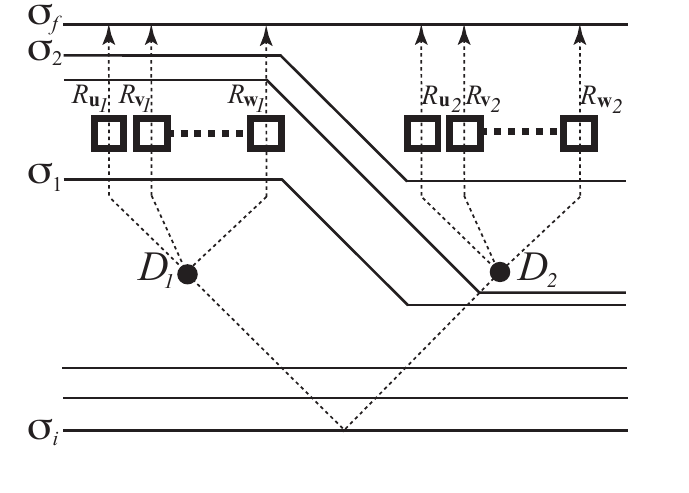}
\caption{\textsf{A Bell test experiment for two entangled isospin particles.
The dashed lines are the (classical) particle trajectories where particle {\it 1} moves initially
to the left and particle {\it 2} moves initially to the right. The vertical represents a timelike direction 
whilst the horizontal represents a spacelike direction. At $D_{\it 1}$ a device is used to deflect
particle {\it 1} towards one of several measuring devices each set up to perform an isospin measurement 
for a different orientation in isospin space. Spacetime regions $R_{{\bf u}_{\it 1}}$, $R_{{\bf v}_{\it 1}}$,$\ldots$,$R_{{\bf w}_{\it 1}}$
are the different interaction regions corresponding to the different isospin orientations
${\bf u}_{\it 1}$, ${\bf v}_{\it 1}$,$\ldots$,${\bf w}_{\it 1}$. Similarly for particle {\it 2}. The state advances through 
a sequence of spacelike surfaces (bold lines) defining a foliation of spacetime. The example foliation shows particle~${\it 1}$ 
measured before particle~${\it 2}$.
}}
\end{center}
\end{figure}

We now suppose that the experimenters at each wing of the apparatus can choose the orientation of their
isospin measurement in isospin space. We suppose that each wing of the experiment now consists of 
several measuring devices each set up to measure the isospin value for different isospin orientations (see figure 3). Each 
particle passes through a deflection device, sending it towards any one of these isospin 
measuring devices. The deflection device can be controlled by the experimenter and each experimenter 
makes their choice of which isospin orientation to measure independently of the other. Furthermore, 
the deflection and measuring devices on one wing of the experiment are completely spacelike separated from the 
deflection and measuring devices on the other wing. This is essentially the experimental design used by 
Aspect in his tests of Bell inequalities \cite{aspe}.

We can represent the initial singlet state in terms of isospin eigenstates in a basis defined by the arbitrarily 
chosen measurement directions. Suppose that the chosen measurement directions correspond to the unit isospin 
vectors ${\bf n}_{\it 1}$ and ${\bf n}_{\it 2}$ and that the angle between ${\bf n}_{\it 1}$ and ${\bf n}_{\it 2}$ is $\theta$, then
\bea
|\psi(\sigma_i)\rangle &=& \sqrtwo\left\{
\cos\left(\thetwo\right) |+\half\rangle_{{\bf n}_{\it 1}}|-\half\rangle_{{\bf n}_{\it 2}}
- i \sin\left(\thetwo\right)|+\half\rangle_{{\bf n}_{\it 1}}|+\half\rangle_{{\bf n}_{\it 2}} \right.
\nonumber\\
&& \;\;\;\;\;\; \left.+ i \sin\left(\thetwo\right)|-\half\rangle_{{\bf n}_{\it 1}}|-\half\rangle_{{\bf n}_{\it 2}}
-\cos\left(\thetwo\right)|-\half\rangle_{{\bf n}_{\it 1}}|+\half\rangle_{{\bf n}_{\it 2}}
\right\},
\eea
where, for isospin vector operators ${\bf S}_a$, the orthonormal eigenstates satisfy
\bea
{\bf n}_{a} \cdot {\bf S}_a|+\half\rangle_{{\bf n}_a} = \half|+\half\rangle_{{\bf n}_a} \;\; ; \;\;
{\bf n}_{a} \cdot {\bf S}_a|-\half\rangle_{{\bf n}_a} = -\half|-\half\rangle_{{\bf n}_a}.
\eea

We denote the spacetime locations of the deflection devices as $D_a$ and the particle-measuring device 
interaction regions as $R_{{\bf u}_a}$, $R_{{\bf v}_a}$,$\ldots$,$R_{{\bf w}_a}$ for the different 
measurement directions ${\bf u}_a$, ${\bf v}_a$,$\ldots$,${\bf w}_a$ (see figure 3). For each $a$, a choice of measurement direction
${\bf n}_{a}$ is made and only one interaction region $R_{{\bf n}_a}$ is activated.
Given ${\bf n}_{\it 1}$ and ${\bf n}_{\it 2}$, the equations of motion for the state are now
\bea
\rd_x |\psi(\sigma)\rangle &=& \left\{ 2\lambda{\bf n}_{\it 1} \cdot {\bf S}_{\it 1} \rd \xi_x^{\it 1}  
-\half \lambda^2 \rd\omega  \right\}|\psi(\sigma)\rangle\;\; {\rm for} \;\; x\in R_{{\bf n}_{\it 1}},
\nonumber\\
\rd_x |\psi(\sigma)\rangle &=& \left\{ 2\lambda {\bf n}_{\it 2} \cdot {\bf S}_{\it 2} \rd \xi_x^{\it 2} 
-\half \lambda^2 \rd\omega  \right\}|\psi(\sigma)\rangle\;\; {\rm for} \;\; x\in R_{{\bf n}_{\it 2}},
\nonumber\\
\rd_x |\psi(\sigma)\rangle &=& 0 \;\; {\rm otherwise},
\label{belEOM}
\eea
where the stochastic increments have the generalised properties
\bea
\rd \xi^a_x &=& 0, \;\; \text{for $x\notin R_{{\bf n}_a}$}; \nonumber\\
\mathbb{E}^{\mathbb{Q}}[\rd \xi^a_x|{\cal F}_{\sigma}^{\xi}] &=& 0, \;\; \text{for $x$ to the future of $\sigma$}; \nonumber\\
\rd \xi^a_x \rd \xi^b_y &=& \delta^{ab}\delta_{xy}\rd\omega, \;\; \text{for $x\in R_{{\bf n}_a}$, $y\in R_{{\bf n}_b}$}.
\label{BrownProps2}
\eea
These equations describe state reduction onto isospin eigenstates defined with respect to the ${\bf n}_{\it 1}$ 
and ${\bf n}_{\it 2}$ directions. Again we consider these equations as effective descriptions of the particle
behaviour resulting from interactions with macroscopic measuring devices.

The solution of (\ref{belEOM}) for an initial isospin singlet state is found to be
\bea
|\psi(\sigma)\rangle &=& \sqrtwo\left\{
\cos\left(\thetwo\right)
e^{\lambda \xi_{\sigma}^{\it 1}-\lambda^2\omega_{\sigma}^{\it 1}} e^{-\lambda \xi_{\sigma}^{\it 2}-\lambda^2\omega_{\sigma}^{\it 2}}
|+\half\rangle_{{\bf n}_{\it 1}}|-\half\rangle_{{\bf n}_{\it 2}}\right.
\nonumber\\&& \;\;\;\;\;\; 
- i \sin\left(\thetwo\right)
e^{\lambda \xi_{\sigma}^{\it 1}-\lambda^2\omega_{\sigma}^{\it 1}} e^{\lambda \xi_{\sigma}^{\it 2}-\lambda^2\omega_{\sigma}^{\it 2}}
|+\half\rangle_{{\bf n}_{\it 1}}|+\half\rangle_{{\bf n}_{\it 2}}
\nonumber\\&& \;\;\;\;\;\;
+ i \sin\left(\thetwo\right)
e^{-\lambda \xi_{\sigma}^{\it 1}-\lambda^2\omega_{\sigma}^{\it 1}} e^{-\lambda \xi_{\sigma}^{\it 2}-\lambda^2\omega_{\sigma}^{\it 2}}
|-\half\rangle_{{\bf n}_{\it 1}}|-\half\rangle_{{\bf n}_{\it 2}}
\nonumber\\&& \;\;\;\;\;\; \left.
-\cos\left(\thetwo\right)
e^{-\lambda \xi_{\sigma}^{\it 1}-\lambda^2\omega_{\sigma}^{\it 1}} e^{\lambda \xi_{\sigma}^{\it 2}-\lambda^2\omega_{\sigma}^{\it 2}}
|-\half\rangle_{{\bf n}_{\it 1}}|+\half\rangle_{{\bf n}_{\it 2}}
\right\}.
\label{BTS}
\eea
As demonstrated in sections \ref{sec:qsol} and \ref{sec:psol} it is straightforward to show that the characteristic 
function associated with the $\mathbb{Q}$-Brownian processes $\xi^{\it 1}_{\sigma}$ and $\xi^{\it 2}_{\sigma}$ (equation (\ref{pchar})) 
can be reproduced directly in the $\mathbb{P}$-measure if we define
\bea
\xi_{\sigma}^{\it 1}&=&4\lambda s_{\it 1}\omega_{\sigma}^{\it 1} + B_{\sigma}^{\it 1},
\nonumber\\
\xi_{\sigma}^{\it 2}&=&4\lambda s_{\it 2}\omega_{\sigma}^{\it 2} + B_{\sigma}^{\it 2},
\eea
where $B^a_{\sigma}$ are $\mathbb{P}$-Brownian motions and the random variables $s_a$ now have the 
joint conditional probability distribution
\bea
\mathbb{P}\left(\left. s_{\it 1} = +\half , s_{\it 2} = -\half \right| {\bf n}_{\it 1}, {\bf n}_{\it 2} \right) 
&=& \half\cos^2\left( \thetwo \right), \nonumber\\
\mathbb{P}\left(\left. s_{\it 1} = +\half , s_{\it 2} = +\half \right| {\bf n}_{\it 1}, {\bf n}_{\it 2} \right) 
&=& \half\sin^2\left( \thetwo \right), \nonumber\\
\mathbb{P}\left(\left. s_{\it 1} = -\half , s_{\it 2} = -\half \right| {\bf n}_{\it 1}, {\bf n}_{\it 2} \right) 
&=& \half\sin^2\left( \thetwo \right), \nonumber\\
\mathbb{P}\left(\left. s_{\it 1} = -\half , s_{\it 2} = +\half \right| {\bf n}_{\it 1}, {\bf n}_{\it 2} \right) 
&=& \half\cos^2\left( \thetwo \right).
\eea

We assume a filtration ${\cal G}_{\sigma}$ such that $B_{\sigma}^{a}$ and $s_a$ are specified.
However, since the probability distribution for $s_{\it 1}$ and $s_{\it 2}$ depends on both experimenters' choice of
measurement directions, we cannot simply assume that $s_a$ are ${\cal G}_{\sigma_i}$-measurable. To
understand the structure of the filtration we can treat the parameters ${\bf n}_{\it 1}$ and ${\bf n}_{\it 2}$ 
as random variables which are independent of any other random variables or processes in the system we 
are describing. We assume that ${\bf n}_{\it 1}$ and ${\bf n}_{\it 2}$ are specified by ${\cal G}_{\sigma}$ in such a 
way that ${\bf n}_a$ is ${\cal G}_{\sigma}$-measurable if and only if the deflection event 
for particle $a$ is to the past of $\sigma$. Note that within this filtration, the variable ${\bf n}_a$ is associated
with the entire surface $\sigma$.  

For a given spacetime foliation the isospin measurement on one wing of the 
apparatus may be complete before the other experimenter has chosen their direction. Suppose for definiteness 
that a given foliation has $R_{{\bf n}_{\it 1}}$ before $D_{\it 2}$ (see figure 3). In order to realise the process 
$\xi^{\it 1}_{\sigma}$ say, it is necessary to realise a definite $s_{\it 1}$. Since ${\bf n}_{\it 2}$ is not 
${\cal G}_{\sigma}$-measurable for spacelike surfaces which have not crossed $D_{\it 2}$, it is 
necessary to show that the marginal distribution of $s_{\it 1}$ is independent of ${\bf n}_{\it 2}$.

In fact we have
\bea
\mathbb{P}\left(\left.s_{\it 1} = +\half \right| {\bf n}_{\it 1}, {\bf n}_{\it 2} \right) &=& 
\mathbb{P}\left(\left. s_{\it 1} = +\half , s_{\it 2} = -\half \right| {\bf n}_{\it 1}, {\bf n}_{\it 2} \right) 
+\mathbb{P}\left(\left. s_{\it 1} = +\half , s_{\it 2} = +\half \right| {\bf n}_{\it 1}, {\bf n}_{\it 1} \right) 
\nonumber\\
&=& \half\cos^2\left( \thetwo \right) +
 \half\sin^2\left( \thetwo \right) 
\nonumber\\
&=&\half,
\eea
as required, and similarly for other marginal probabilities. This enables us to draw values of $s_{\it 1}$ from the 
correct probability distribution without knowledge of ${\bf n}_{\it 2}$ which happens in the future for
the given example foliation. In this case we require that $s_{\it 1}$ is ${\cal G}_{\sigma_1}$-measurable 
for some surface $\sigma_1$ to the past of $R_{{\bf n}_{\it 1}}$ (figure 3).

We can define some other surface $\sigma_2$ that is to the past of $R_{\it 2}$ but to the future of $\sigma_1$ 
and both particle deflection events (see figure 3). Since ${\bf n}_{\it 1}$, ${\bf n}_{\it 2}$, and $s_{\it 1}$, are all 
${\cal G}_{\sigma_2}$-measurable we can write,
for example,
\bea
\mathbb{P}\left(\left. s_{\it 2}=+\half \right| {\cal G}_{\sigma_2} \right) &=&
\mathbb{P}\left(\left. s_{\it 2}=+\half \right| s_{\it 1}=+\half ; {\bf n}_{\it 1}, {\bf n}_{\it 2}\right) 
\nonumber \\ &=& 
\frac{\mathbb{P}\left(\left. s_{\it 1} = +\half , s_{\it 2} = +\half \right| {\bf n}_{\it 1}, {\bf n}_{\it 2} \right) }
{\mathbb{P}\left(\left.s_{\it 1} = +\half \right| {\bf n}_{\it 1}, {\bf n}_{\it 2} \right)}
=\sin^2\left( \thetwo \right),
\eea
and similarly for other conditional probabilities. This enables us to draw values of $s_{\it 2}$ from the 
correct probability distribution with global knowledge of ${\bf n}_{\it 1}$, ${\bf n}_{\it 2}$, and $s_{\it 1}$.
We can therefore say that $s_{\it 2}$ is ${\cal G}_{\sigma_2}$-measurable.

For a different foliation where $R_{{\bf n}_{\it 2}}$ precedes $D_{\it 1}$ we would use the marginal probability 
distribution to determine $s_{\it 2}$ and the conditional distribution to determine $s_{\it 1}$. In any
case the joint distribution is the same. The order in which $s_{\it 1}$ and $s_{\it 2}$ are assigned 
has no physical significance. It is simply related to our arbitrary choice of spacetime foliation
within the covariant Tomonaga picture of state evolution. We also stress that the random variables 
$s_a$ were introduced to facilitate solution of the dynamical equations. They are not part of the 
physical model as originally presented. The purpose of the argument presented here is simply to 
show that the picture of state evolution is consistent and does not require prior knowledge of
the experimenter's decisions.

\subsection{State reduction}

State reduction follows from the solution in the same way as shown in section \ref{SR}. For example,
given ${\bf n}_{\it 1}$ and ${\bf n}_{\it 2}$ we condition on the event $s_{\it 1}= +1/2$, $s_{\it 2} = +1/2$. The unnormalised 
expectation of the spin operator for particle ${\it 1}$ is found from equation (\ref{BTS}) to be
\bea
\langle\psi(\sigma)| {\bf n}_{\it 1}\cdot{\bf S}_{\it 1}|\psi(\sigma)\rangle &=& \half 
e^{2\lambda B_{\sigma}^{\it 1}+2\lambda^2\omega_{\sigma}^{\it 1}}
e^{2\lambda B_{\sigma}^{\it 2}+2\lambda^2\omega_{\sigma}^{\it 2}}
\left\{
\cos^2\left(\thetwo \right)\left(e^{-4\lambda B_{\sigma}^{\it 2}-8\lambda^2\omega_{\sigma}^{\it 2}}
-e^{-4\lambda B_{\sigma}^{\it 1}-8\lambda^2\omega_{\sigma}^{\it 1}}\right) \right. \nonumber\\ 
&& \left. +\sin^2\left(\thetwo\right)\left(1
-e^{-4\lambda B_{\sigma}^{\it 1}-8\lambda^2\omega_{\sigma}^{\it 1}}
e^{-4\lambda B_{\sigma}^{\it 2}-8\lambda^2\omega_{\sigma}^{\it 2}} 
\right)\right\},
\eea
and the state norm is
\bea
\langle\psi(\sigma)|\psi(\sigma)\rangle &=& 
e^{2\lambda B_{\sigma}^{\it 1}+2\lambda^2\omega_{\sigma}^{\it 1}}
e^{2\lambda B_{\sigma}^{\it 2}+2\lambda^2\omega_{\sigma}^{\it 2}}
\left\{
\cos^2\left(\thetwo\right)\left(e^{-4\lambda B_{\sigma}^{\it 2}-8\lambda^2\omega_{\sigma}^{\it 2}}
+e^{-4\lambda B_{\sigma}^{\it 1}-8\lambda^2\omega_{\sigma}^{\it 1}}\right) \right. \nonumber\\ 
&& \left. +\sin^2\left(\thetwo\right)\left(1
+e^{-4\lambda B_{\sigma}^{\it 1}-8\lambda^2\omega_{\sigma}^{\it 1} }
e^{-4\lambda B_{\sigma}^{\it 2}-8\lambda^2\omega_{\sigma}^{\it 2}} 
\right)\right\}.
\eea
Using equation (\ref{problimit}) we then find that as $\omega^{\it 1}_{\sigma}\rightarrow\infty$,
\bea
\langle {\bf n}_{\it 1}\cdot{\bf S}_{\it 1}\rangle_{\sigma}  = \frac{\langle \psi(\sigma)|{\bf n}_{\it 1}\cdot{\bf S}_{\it 1} |\psi(\sigma)\rangle}
{\langle \psi(\sigma) | \psi(\sigma)\rangle}\rightarrow\half.
\eea
As expected the isospin of particle ${\it 1}$ in the direction ${\bf n}_{\it 1}$ tends to the value $\half$. A similar 
calculation shows that $\langle {\bf n}_{\it 2}\cdot{\bf S}_{\it 2}\rangle_{\sigma}  \rightarrow \half$
as $\omega^{\it 2}_{\sigma}\rightarrow\infty$, along with 
similar results for other given values of~$s_a$. 

It is also straightforward to show that 
\bea
\lim_{\omega^{\it 1}_{\sigma},\omega^{\it 2}_{\sigma}\rightarrow\infty}\langle 
({\bf n}_{\it 1}\cdot{\bf S}_{\it 1}) ({\bf n}_{\it 2}\cdot{\bf S}_{\it 2})\rangle_{\sigma}=
\left\{\begin{array}{rl}
\quat & {\rm with \; probability \;} \sin^2\left(\thetwo\right), \\ \\
-\quat & {\rm with \; probability \;} \cos^2\left(\thetwo\right), 
\end{array}\right.
\eea
such that 
\bea
\mathbb{E}^{\mathbb{P}} \left[\left. \lim_{\omega^{\it 1}_{\sigma},\omega^{\it 2}_{\sigma}\rightarrow\infty}
\langle ({\bf n}_{\it 1}\cdot{\bf S}_{\it 1}) ({\bf n}_{\it 2}\cdot{\bf S}_{\it 2})\rangle_{\sigma} 
\right| {\cal F}_{\sigma_i}^{\xi}\right]
= -\quat \cos \theta = -\quat {\bf n}_{\it 1}\cdot {\bf n}_{\it 2}.
\eea
This agrees with the result predicted by standard quantum theory and is confirmed by Bell test experiments.

\subsection{Parameter independence}

The parameter independence condition states that the probability of a given outcome for an isospin 
measurement on one wing of the experiment is independent of the chosen measurement direction on the 
other wing. This is an important feature since if the model were parameter dependent we could transmit 
messages at superluminal speeds.

Parameter independence can be stated as follows:
\bea
\mathbb{P}\left(\left. \lim_{\omega^{\it 1}_{\sigma}\rightarrow\infty}\langle {\bf n}_{\it 1}\cdot {\bf S}_{\it 1}\rangle_{\sigma}
=+ \half \right| {\cal F}_{\sigma_i}^{\xi}; {\bf n}_{\it 1},{\bf n}_{\it 2} \right) = 
\mathbb{P}\left(\left. \lim_{\omega^{\it 1}_{\sigma}\rightarrow\infty}\langle {\bf n}_{\it 1}\cdot {\bf S}_{\it 1}\rangle_{\sigma}
=+ \half \right| {\cal F}_{\sigma_i}^{\xi}; {\bf n}_{\it 1} \right),
\label{paramIndep2}
\eea
and similarly for $1 \leftrightarrow 2$. In order to prove this relation we define projection operators
$P^{+}_{{\bf n}_a}$ by
\bea
P^+_{{\bf n}_a}|+\half \rangle_{{\bf n}_a} = |+\half \rangle_{{\bf n}_a} \;\; ; \;\;
P^+_{{\bf n}_a}|-\half \rangle_{{\bf n}_a} = 0.
\eea
In the limit that $\omega^{\it 1}_{\sigma}\rightarrow\infty$ we can write
\bea
\mathbb{P}\left(\left. \langle {\bf n}_{\it 1}\cdot {\bf S}_{\it 1}\rangle_{\sigma}
=+ \half \right| {\cal F}_{\sigma_i}^{\xi}; {\bf n}_{\it 1},{\bf n}_{\it 2} \right) &=& 
\mathbb{E}^{\mathbb{P}}\left[\left.\langle P^+_{{\bf n}_{\it 1}} \rangle_{\sigma} 
\right| {\cal F}_{\sigma_i}^{\xi}; {\bf n}_{\it 1},{\bf n}_{\it 2} \right]\nonumber\\
&=&
\mathbb{E}^{\mathbb{Q}}\left[\left. \langle\psi(\sigma)| P^+_{{\bf n}_{\it 1}}|\psi(\sigma)\rangle \right| 
{\cal F}_{\sigma_i}^{\xi}; {\bf n}_{\it 1},{\bf n}_{\it 2} \right] 
\nonumber\\
&=&\half\mathbb{E}^{\mathbb{Q}}\left[\left.
\cos^2\left( \thetwo \right)
e^{2\lambda \xi_{\sigma}^{\it 1}-2\lambda^2\omega_{\sigma}^{\it 1}}
e^{-2\lambda \xi_{\sigma}^{\it 2}-2\lambda^2\omega_{\sigma}^{\it 2}}
\right| {\cal F}_{\sigma_i}^{\xi}; {\bf n}_{\it 1},{\bf n}_{\it 2} \right]
\nonumber\\ 
&&+
\half\mathbb{E}^{\mathbb{Q}}\left[\left.
\sin^2\left( \thetwo \right)
e^{2\lambda \xi_{\sigma}^{\it 1}-2\lambda^2\omega_{\sigma}^{\it 1}}
e^{2\lambda \xi_{\sigma}^{\it 2}-2\lambda^2\omega_{\sigma}^{\it 2}}
\right| {\cal F}_{\sigma_i}^{\xi}; {\bf n}_{\it 1},{\bf n}_{\it 2} \right]
\nonumber\\
&=& \half\cos^2\left( \thetwo \right)
+\half\sin^2\left( \thetwo \right)
\nonumber\\
&=& \half.
\eea
The probability of a given outcome for particle {\it 1} is independent of ${\bf n}_{\it 2}$ as required.

\section{The free will theorem}
\label{sec:freewill}

The Free Will Theorem of Conway and Kochen \cite{fw1,fw2} asserts that if an experimenter is free to 
make decisions about which directions to orient their apparatus in a spin measurement, then the response 
of the spin particle cannot be a function of information content in the part of the universe 
that is earlier than the response itself. The conclusion of Conway and Kochen is that this rules out the 
possibility of being able to formulate a relativistic model of dynamical state reduction. It is claimed 
that a classical stochastic process which dictates a definite spin measurement outcome must be considered
to be information as defined within the theorem. The theorem then states that the particle's response 
cannot be determined by this classical information, undermining the construction of dynamical models 
of state reduction. We do not reproduce the proof of the theorem here (it can be found in \cite{fw1,fw2}). 
In order to understand that the conclusion of Conway and Kochen is inappropriate it will suffice to analyse 
the three axioms of the Free Will Theorem with reference to the model outlined in this paper.

The first axiom SPIN specifies the existence of a spin-1 particle for which measurements of the squared 
components of spin performed in three orthogonal directions will always yield the results 1, 0, 1 in some 
order. The second axiom TWIN asserts that it is possible to form an entangled pair of spin-1 particles in a 
combined singlet state such that if measurements of the components of squared spin were performed in the same 
direction for each particle they would yield identical results. These two axioms follow directly from the 
quantum mechanics of spin particles. A situation is considered where experimenters at spacelike separated 
locations $D_{\it 1}$ and $D_{\it 2}$ can each choose the orthogonal set of directions in which to measure 
the components of squared spin for each particle. (The proof of the Free Will Theorem makes use of the Peres 
configuration of 33 directions for which it can be shown that it is impossible to find a function on the set of directions
with the property that its value for any orthogonal set of directions is always 1, 0, 1 in some order.)
Although we have considered a different spin system in this paper, the similarities between the experimental 
set-ups allow us to evaluate the applicability of the Free Will Theorem to dynamical state reduction. 

The third axiom MIN (in the latest version of the proof \cite{fw2}) states that the particle response at 
$R_{{\bf n}_{\it 1}}$ (using our notation where it is understood that the choice of spin measurement direction 
${\bf n}_{\it 1}$ corresponds to an orthogonal triple of directions) is independent of the choice of measurement 
direction at $D_{\it 2}$ and similarly that the particle response at $R_{{\bf n}_{\it 2}}$ is independent of the 
choice of measurement direction at $D_{\it 1}$. Information is defined in the context of MIN in such a way that 
any information which influences the measurement outcome at $R_{{\bf n}_{\it 1}}$ is independent of ${\bf n}_{\it 2}$ 
and any information which influences the measurement outcome at $R_{{\bf n}_{\it 2}}$ is independent of ${\bf n}_{\it 1}$. 
We can immediately see that this definition of information does not apply to the classical stochastic processes
$\xi_{\sigma}^a$ considered in our model. As highlighted above, $\xi_{\sigma}^a$ can be expressed in terms 
of a random variable $s_a$ whose value corresponds to the eventual spin measurement outcome, and a physical
Brownian motion process $B_{\sigma}^a$ which acts as a noise term, obscuring the value of $s_a$. The 
realised value of $s_a$ indeed depends on the choice of measurement direction at the opposite wing of the 
experiment in the way shown in section \ref{sec:bell}. Since the process $\xi_{\sigma}^a$ influences the 
measurement outcome in a way which depends critically on the realised value of $s_a$, it does not satisfy 
the definition of MIN information. Furthermore, there is no reason why the mechanism of state reduction 
outlined in this paper cannot be applied to any spin system including the TWIN SPIN system used to
prove the Free Will Theorem.

More generally we are able to see that the MIN axiom need not be satisfied whilst still maintaining independence 
from any specific inertial frame. Viewing state evolution in the Tomonaga picture we must choose a foliation 
of spacetime to provide a framework for a consistent narrative of the state 
evolution. Covariance enters with the fact that all choices of foliation are equivalent; the state can be 
defined on any spacelike hypersurface. For a foliation where $R_{{\bf n}_{\it 1}}$ happens before $D_{\it 2}$, the state 
will collapse across the entire hypersurface as it crosses $R_{{\bf n}_{\it 1}}$, to a new state consistent with 
the isospin measurement direction ${\bf n}_{\it 1}$. In this way the response of particle ${\it 1}$ is independent of the choice of 
measurement direction at $D_{\it 2}$ (which happens later in the evolution) but the response of particle~${\it 2}$ depends 
(via the collapsed state) on the random variable $\theta$. The opposite interpretation can be made for a 
foliation where $R_{{\bf n}_{\it 2}}$ is before $D_{\it 1}$. Thus the MIN axiom should read that {\it either} the particle 
response at $R_{{\bf n}_{\it 2}}$ is independent of the choice of measurement direction at $D_{\it 1}$ {\it or} the particle 
response at $R_{{\bf n}_{\it 1}}$ is independent of the choice of measurement direction at $D_{\it 2}$, the difference 
being a matter of interpretation. With this modification the proof of the Free Will Theorem no longer holds.  

We stress that the choice of spacetime foliation is analogous to an arbitrary gauge choice. It allows us to form a 
global covariant picture of state evolution without reference to any individual observer's frame.

\section{Conclusions}

We have argued that the principles of quantum mechanics are in need of modification 
if we hope to find a unified description of micro and macro behaviour. We have seen that 
alternatives to quantum dynamics can feasibly be constructed despite the apparent 
invulnerability of standard quantum theory when faced with experimental evidence. It may even be possible 
to test new theories against standard quantum theory in the near future~\cite{pearexpt, legg}.
 
We have demonstrated a continuous state reduction dynamics describing the measurement of two 
spacelike separated spin particles in an EPR experiment. The correlation between measured
outcomes for the two particles, particularly when the experimenters are free to choose the 
orientations of their spin measurements, offers an interesting challenge for dynamical models 
of state reduction. We have seen that the use of the physical probability measure induces a
corresponding correlation between the stochastic processes to which the particle states are
coupled. State evolution is covariantly described using the Tomonaga picture with no dependence
on any chosen frame and no possibility for superluminal communication. The results of measurements 
agree with standard quantum theory, in particular for the purpose of performing a test of Bell 
inequalities for the system.

The value of this model is to show that the state reduction process can indeed be described
by a relativistically-invariant stochastic dynamics (contrary to the claims of Conway and Kochen). 
We have shown how to solve the dynamical equations and this has led to new insight into the 
structure of the filtration. In the physical measure, the covariantly-defined stochastic 
processes are seen to be constructed from a random variable which relates directly to the 
measurement outcome and a noise process which obscures the random variable, making it inaccessible
from the point of view of the state dynamics. This allows us to reinterpret the problem of solving
the stochastic equations of motion as a nonlinear filtering problem whereby the aim is to form 
a best estimate of the hidden random variable based only on information contained in the 
observable processes. It is hoped that these insights might help to indicate ways in which we 
might tackle state reduction dynamics in relativistic quantum field systems.

\section*{Acknowledgements}
I would like to thank Dorje Brody and Lane Hughston for a series of useful discussion sessions.
I would also like to thank the Theoretical Physics Group at Imperial College where this work 
was carried out.

\end{document}